\documentclass[sigconf, nonacm]{acmart}

\AtBeginDocument{%
  \providecommand\BibTeX{{%
    \normalfont B\kern-0.5em{\scshape i\kern-0.25em b}\kern-0.8em\TeX}}}

\setcopyright{acmcopyright}
\copyrightyear{2018}
\acmYear{2018}
\acmDOI{XXXXXXX.XXXXXXX}

\acmConference[Conference acronym 'XX]{Make sure to enter the correct
  conference title from your rights confirmation emai}{June 03--05,
  2018}{Woodstock, NY}
\acmPrice{15.00}
\acmISBN{978-1-4503-XXXX-X/18/06}

\usepackage{xspace}
\usepackage{subcaption}
\graphicspath{ {./images/} }

\newcommand{\one}{({\em i}\/)\xspace}
\newcommand{\two}{({\em ii}\/)\xspace}
\newcommand{\three}{({\em iii}\/)\xspace}

\def\eg{\emph{e.g.}\xspace}
\def\ie{\emph{i.e.}\xspace}

\newcommand{\pb}[1]{\vspace{0.75ex}\noindent{\bf \em #1}\hspace*{.3em}}

\begin{document}

\title{Collaborative Content Moderation in the Fediverse}

\author{Haris Bin Zia}
\email{h.b.zia@qmul.ac.uk}
\affiliation{%
  \institution{QMUL}
  \city{London}
  \country{United Kingdom}
}

\author{Aravindh Raman}
\email{araram@cisco.com}
\affiliation{%
  \institution{Cisco ThousandEyes}
  \city{London}
  \country{United Kingdom}
}

\author{Ignacio Castro}
\email{i.castro@qmul.ac.uk}
\affiliation{%
  \institution{QMUL}
  \city{London}
  \country{United Kingdom}
}

\author{Gareth Tyson}
\authornote{Also with QMUL}
\email{gtyson@ust.hk}
\affiliation{%
  \institution{HKUST (GZ)}
  \city{Guangzhou}
  \country{China}
}

\renewcommand{\shortauthors}{Zia et al.}

\begin{abstract}

The Fediverse, a group of interconnected servers providing a variety of interoperable services (\eg micro-blogging in Mastodon) has gained rapid popularity.
This sudden growth, partly driven by Elon Musk's acquisition of Twitter, has created challenges for administrators though.
This paper focuses on one particular challenge: content moderation, \eg the need to remove spam or hate speech.
While centralized platforms like Facebook and Twitter rely on automated tools for moderation, their dependence on massive labeled datasets and specialized infrastructure renders them impractical for decentralized, low-resource settings like the Fediverse. 
In this work, we design and evaluate \textit{FedMod}, a collaborative content moderation system based on federated learning. Our system enables servers to exchange parameters of partially trained local content moderation models with similar servers, creating a federated model shared among collaborating servers. FedMod demonstrates robust performance on three different content moderation tasks: harmful content detection, bot content detection, and content warning assignment, achieving average per-server macro-F1 scores of 0.71, 0.73, and 0.58, respectively.

\end{abstract}

\maketitle

\section{Introduction}
\label{sec:intro}

The proliferation of social media has reshaped the landscape of online communication, enabling individuals to connect, share, and engage with a global audience. As these platforms have grown in popularity, so has the challenge of \emph{content moderation}~\cite{singhal2022sok}.
Content moderation refers to the process of reviewing and regulating user-generated content to uphold platform policies and community standards. While traditional `centralized' platforms like Twitter and Facebook have long grappled 
with content moderation challenges~\cite{gillespie2020content,del2017hate}, the advent of the Fediverse --- a decentralized network of interconnected social media servers --- has introduced a unique set of new complexities to the content moderation landscape~\cite{anaobi2023will,bin2022toxicity}. 

The Fediverse is an ensemble of interconnected platform servers 
used for micro-blogging (\eg Mastodon~\cite{mastodonurl} \& Pleroma~\cite{pleromaurl}), social networking (\eg Diaspora~\cite{diasporaurl}), and video sharing (\eg PeerTube~\cite{peertubeurl}).
Several unique features differentiate these platforms from their centralized counterparts: 
\one~They consist of independently owned, operated, and moderated communities located on independent lightweight servers known as ``instances''; 
\two~They enable users to own their data --- in fact, some users even choose to fork their own instance to keep complete control over their data; 
\three~They allow users to interact locally (within instances) as well as globally (across instances) via so-called \emph{federation} --- this involves instances interconnecting in a peer-to-peer fashion, allowing their users to communicate.

This decentralized architecture presents unique challenges for content moderation.
Typical centralized social networking platforms rely on centrally trained large-scale models for automated content moderation.
For instance, Facebook  relies on machine learning  to identify 97\% of the content that the platform removes for violating its hate speech policies.\footnote{https://about.fb.com/news/2021/02/update-on-our-progress-on-ai-and-hate-speech-detection}
This is infeasible in the Fediverse though. Fediverse administrators are often volunteers with limited time and computational resources who only control the data within their specific instances. 
Thus, manual content moderation often falls upon volunteers, who can quickly become overwhelmed~\cite{hassan2021exploring}. %
We argue that addressing this issue is crucial to ensure the success of Fediverse applications.

To overcome the above challenges, we explore the potential of automated content moderation within the Fediverse. Our study centers on the largest decentralized micro-blogging social network, Mastodon. We collect data from 50 unique Mastodon instances, comprising 219,577 unique posts from 44,840 users. Our work examines three key content moderation tasks. 
First, we investigate the potential for automated tools in identifying ``toxic'' or ``harmful'' material. To create a ground truth dataset, we utilize the Perspective API~\cite{perspective} to label all Mastodon posts, serving as a proxy for the actual labels from instance administrators. 
However, a natural limitation of the above is that it creates a single consistent label definition (as dictated by Perspective). In a real scenario, each instance administrator would have to label their own posts, potentially leading to subjectivity in what is considered ``toxic''.
Thus, to better understand the challenges of decentralized labeling, we study a second task using a dataset of ``content warning'' posts instead. These are posts that are self-labeled by their authors. These are not necessarily toxic, but nevertheless may offend some users. Importantly, because they are self-labeled, we can study the impact of inconsistent and subjective training labels.
The above two tasks focus on per-post issues. To complement these tasks, we finally look at the detection of bot accounts. Therefore, the third task attempts to automatically detect bot posts, using a dataset of posts from self-declared Mastodon bot accounts.

With these tasks in-mind, we first explore the potential of local content moderation in each individual instance.
We assess the ability of administrators to train their own local classification models for automatically moderating content (\S\ref{sec:potentialoflocalcontentmoderation}).
This involves either each administrator annotating a set of their local instance's posts (\eg for harmful content detection) or utilizing crowd-sourced labels from the users of the instances (\eg for bot content detection or content warning assignment) and training their own classifier. %
We confirm that it \textit{is} indeed possible to build such local models. The local models achieve an average macro-F1 score of 0.63 for harmful content detection and 0.69 for bot content detection.
However, the performance is quite poor for content warning assignment, with a macro-F1 score of 0.53. Furthermore, the performance of these local models varies significantly across instances.
For example, for harmful content detection, \texttt{social.eiden.ch}'s local model attains a macro-F1 score of 0.76, whereas \texttt{cryptodon.chat} attains a score of just 0.36.
To make matters worse, these locally trained models place a large burden on administrators who might need to label thousands of posts. We argue that these problems emerge from a missed opportunity: although instances share posts via federation, they do not share content moderation knowledge.

To exploit this observation, we propose \textit{FedMod}, a collaborative content moderation system based on federated learning~\cite{li2020federated}.
In FedMod, instances exchange parameters of partially trained local models to assist each other in a peer-to-peer fashion (\S\ref{sec:collaborativecontentmoderation}). 
Collaborative moderation offers the advantage of implicitly sharing knowledge %
across instances while preserving privacy by not requiring administrators to disclose their local labeled posts. %
FedMod operates a multi-step process. 
First, each instance utilizes its own limited set of labeled posts to build a local content moderation model (\eg to classify harmful posts). 
Then, each instance searches for peers that it considers sufficiently `similar' to collaborate with. 
Finally, the instances employ federated learning to exchange parameters of their local model between each other. Iteratively, this process allows instances to co-train moderation models.
We demonstrate that compared to local models, FedMod improves automated content moderation for all three tasks across all tested instances. It improves harmful content detection by 12.69\%, bot content detection by 5.79\%, and content warning assignment by 9.43\%, resulting in average per-instance macro-F1 scores of 0.71, 0.73, and 0.58, respectively.
Although we test FedMod on three content moderation tasks, our approach can also be applied to other classification tasks (\eg spam detection).

\section{Background \& Challenges}
\label{sec:background}

\subsection{A Primer on the Fediverse}

\pb{Fediverse.}
The Fediverse is a network of independently owned, operated, moderated, and interconnected ``Decentralised Web'' servers (known as ``instances''). 
The architecture of the Fediverse is such that no single entity operates the entire infrastructure. Instead, instances collaborate (aka ``federate'') in a peer-to-peer fashion to collectively offer various types of services (\eg micro-blogging, file sharing, video streaming). This federation is performed using the W3C ActivityPub~\cite{activitypub} protocol, which allows instances to subscribe to objects provided by each other. The nature of these objects varies based on the specific application in question. For example, whereas Mastodon (a micro-blogging platform) exchanges posts, PeerTube (a video-sharing platform) exchanges video information. This allows these instances to form a decentralized network of content exchange.

\pb{Mastodon.}
The largest Fediverse platform is Mastodon,\footnote{https://the-federation.info/\#projects}  an open-source micro-blogging service. 
Each unique Mastodon instance works much like Twitter, allowing users to register new accounts and share \textbf{posts} with their followers --- equivalent to a tweet on Twitter. Users can also \textbf{repost} others' posts --- equivalent to retweet on Twitter.
Instances can work in isolation, only allowing locally registered users to follow each other. 
However, Mastodon instances can also \textbf{federate}, whereby users registered on one instance can follow users registered on another instance. %
This results in the instance \textbf{subscribing} to posts performed on the remote instance, such that they can be pushed across and presented to local users.
We refer to users registered on the same instance as \textbf{local}, and users registered on different instances as \textbf{remote}.

\subsection{Challenges of Content Moderation}
\label{subsec:challengesofcontentmoderation}

Administration on Mastodon is decentralized: each instance administrator(s) formulates and enforces their own local moderation policy. 
This is a radical departure from prior centralized approaches to moderation, where a single administrative entity has full control (\eg Twitter).
Adding further complexity, the administrator of an instance has limited control over the remote content.
This is a fundamental difference with platforms such as Reddit: In Mastodon, posts generated on one instance can easily spread to another instance, even if those two instances have wildly different moderation policies. 
A recent example of this occurred when the \texttt{gab.social} instance joined the Fediverse, rapidly spreading hate speech to other instances~\cite{gabhate}.  
We argue that these challenges together with the large scale of the Fediverse and the limited set    of voluntary administrators, make manual moderation infeasible. 
We believe that addressing this problem is vital to ensure the success of Mastodon and other Fediverse applications.

\section{Data \& Motivation}

\subsection{Dataset}
\label{sec:dataset}

To underpin our work, we gather a large Mastodon dataset.
Our data collection process involves three main steps: \one~Discovering a set of Mastodon instances; \two~Collecting the posts from each instance; and \three~Labeling posts for content moderation tasks.
See Appendix \ref{sec:app:ethics} for our ethics discussion.

\subsubsection{Discovering Instances.} We crawl a list of Mastodon instances from the Fediverse Observer,\footnote{https://fediverse.observer} a comprehensive index that collects statistics about Fediverse instances through both manual registration and automated web crawling. This yields a dataset of 11,516 unique instances, housing a total of 6,825,677 users, with an average of 1,162 users per instance. Recent research suggests that users on Mastodon instances of different sizes (w.r.t. number of users) exhibit distinct behaviors in terms of posting frequency~\cite{zia2023flocking}. Thus, to capture a diverse range of users, we sample 25 random instances with a user count above the average and another random 25 with a user count below the average, totaling 50 instances for our analysis. These 50 instances collectively have 2,437,140 registered users. We limit ourselves to this set, as it would be infeasible to annotate posts for all 11,516 servers.

\subsubsection{Collecting Posts.} 
\begin{figure}[t]
  \centering
  \includegraphics[width=\linewidth]{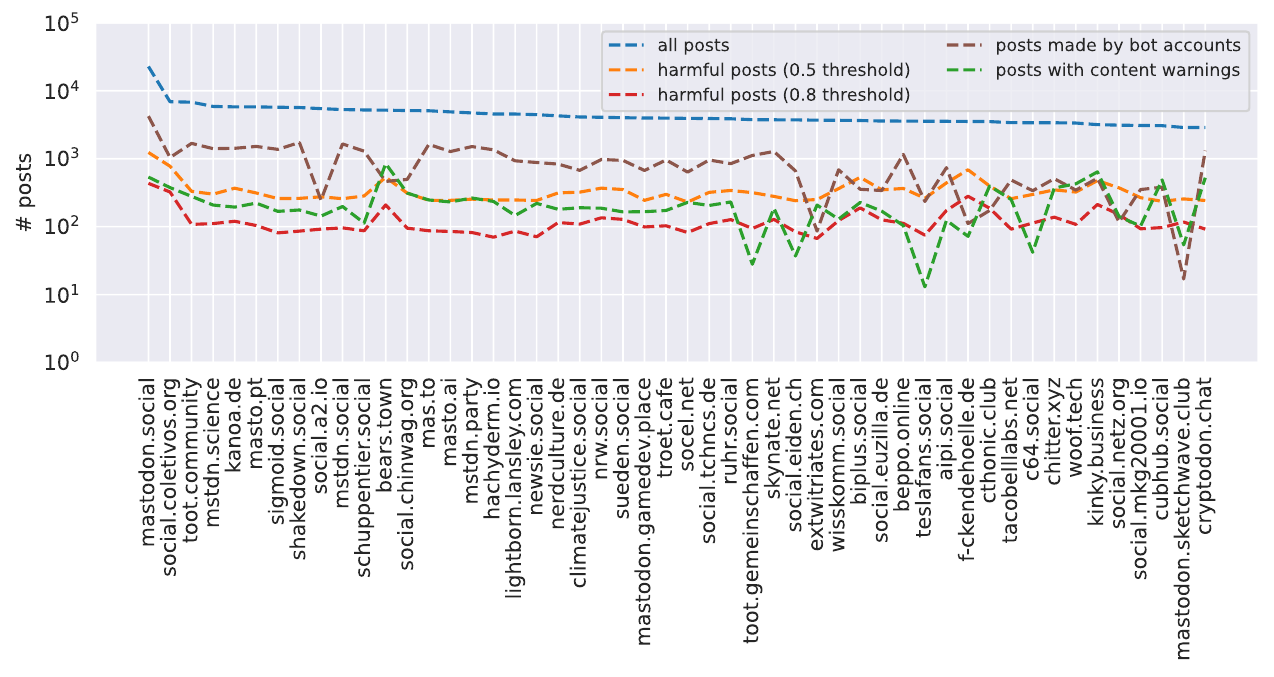}
  \caption{Distribution of number of posts on each instance. Note the log scale on the Y-axis.} %
  \label{lab:dataset}
\end{figure}

Next, we retrieve all public posts from the sampled instances using their Public Timeline API.\footnote{<instance.uri>/api/v1/timelines/public} This endpoint returns all public posts on the instance, along with their associated metadata, including number of likes, number of reports, language, and timestamp. In total, we collect 219,577 unique posts authored by 44,840 unique users between March 01, 2023, and June 21, 2023. We report the number of posts on each instance in Figure~\ref{lab:dataset}. These posts encompass 17 different languages, with English being the most predominant at 75.29\%, followed by German at 17.73\%. %

\subsubsection{Labeling Posts.} We next label Mastodon posts for three content moderation tasks: 

\pb{Harmful Content Detection.} We use the Perspective API~\cite{perspective} to annotate the Mastodon posts as harmful vs.\ non-harmful.\footnote{Refer to the Appendix~\ref{sec:directperspective} for a discussion on why directly using the Perspective API for content moderation in the Fediverse is not a sustainable solution.}
We acknowledge that human labeling by real administrators would be preferable~\cite{perspectivelimits}, but this information is not available.
Note, the Perspective API has undergone extensive evaluations across various social media platforms~\cite{papasavva2021qoincidence,papasavva2020raiders,rottger2020hatecheck} , demonstrating its effectiveness in identifying harmful content.
The Perspective API predicts the perceived impact of each post on a conversation and provides a score ranging from 0 to 1, with 1 being the most harmful (and 0 the least).
This reflects six different emotional attributes, ranging from \texttt{TOXICITY} to \texttt{THREAT} 
(for the definition of each attribute see Appendix~\ref{sec:appendixdataset}). 
To capture a broad range of problematic content, we regard \emph{harmful} posts as those where their Perspective score for any of the attributes is greater than 0.5 (and as non-harmful if it falls below this threshold).
This threshold (0.5) is a common choice in the literature~\cite{papasavva2021qoincidence,papasavva2020raiders,rottger2020hatecheck}.
For completeness, we also repeat experiments with a stricter 0.8 threshold~\cite{kumar2023understanding}, and find the results are similar (see  Appendix~\ref{sec:appendixadditionalevaluation}). One limitation of this labeling methodology is that each instance's annotations are based on a consistent and standardized definition, as determined by the Perspective API. However, in reality, administrators may make mistakes during the annotation process or simply not exhibit consistent views. To address this concern, we later introduce randomized noise into the labels on a per-instance basis to account for discrepancies in annotations (\S\ref{sec:potentialoflocalcontentmoderation}).

\pb{Content Warning Assignment.}
Content warnings are special (visible) tags that accompany posts that are judged to be sensitive (\eg NSFW). These are set by the post author themselves, and are used to warn future viewers. Note, this is different to Perspective labels that indicate toxicity (\ie the presence of a warning does not necessarily imply toxicity).
Whether a post is marked sensitive or not can be checked from the boolean ``sensitive'' parameter in its metadata. We utilize these self-tagged, crowd-sourced labels for content warning assignment. Overall, 11,074 unique posts are self-tagged with content warnings by 3,905 unique users (post authors) across 50 Mastodon instances. %

\pb{Bot Content Detection.} Mastodon users can declare themselves as bots during profile creation, but not all do, making detection challenging. This information is accessible in post metadata as a boolean parameter indicating whether the post is made by a bot account. We use these self-declared, crowd-sourced labels for bot content detection. In our dataset, 42,566 unique posts are made by 3,371 unique bot accounts across 50 Mastodon instances. %

\subsection{Quantifying Need for Automated Moderation}

To motivate our work, we begin by validating the presence of harmful content on Mastodon. 
Figure~\ref{lab:dataset} shows the distribution of harmful posts, the distribution of posts made by bot accounts, as well as the distribution of posts tagged with content warnings, across all instances. 

Overall, 7.48\% of all posts on Mastodon are classified as harmful, with \texttt{f-ckendehoelle.de} having the highest proportion (19.43\%) when using a 0.5 threshold with the Perspective API.
With a threshold of 0.8, this value is 2.71\%, and \texttt{f-ckendehoelle.de} still has the highest percentage of harmful posts (7.92\%).
This underscores the importance of automated content moderation because, in the absence of built-in moderation tools, administrators are compelled to manually moderate such content. 

Furthermore, 19.38\% of all posts are made by self-declared bot accounts, with \texttt{cryptodon.chat} having the highest proportion (45.65\%). Despite 13 out of 50 instances in our dataset explicitly disallowing bots in their content moderation policy (usually listed on the instance's about page), all instances contain posts made by bot accounts. This further emphasizes the need for automated moderation. 
Moreover, 5.04\% of all posts are self-tagged with content warnings, with \texttt{kinky.business} having the highest proportion (20.16\%). Posts tagged with content warnings serve to alert users about potentially sensitive material, allowing them to make informed decisions about their interaction with the content. However, manual (voluntary) tagging by users can be inconsistent, exposing users to unexpected or triggering material. This inconsistency again underscores the necessity for automated tools that ensure the consistent application of content warnings across instances.

\section{Potential of Local Models}
\label{sec:potentialoflocalcontentmoderation}

One intuitive solution is for each administrator to train their own local content moderation model.

\pb{Training \& Testing Sets.}
To test the efficacy of this approach, we first train local content moderation models separately for each Mastodon instance using their respective labeled posts for each content moderation task. For each instance and each task, we perform an 80:20 split of posts, stratified by their respective labels, between the training and testing sets.
However, considering the voluntary nature of instance administrators, we do not utilize the entire training set --- it would be infeasible for a volunteer administrator to annotate hundreds or thousands of posts each day.
Instead, we adopt an online learning approach~\cite{onlinelearning}, using only a small number of labeled posts at each training step. 
Specifically, we train models for 8 steps using a randomly sampled set of 16 posts from the training set at each step.
We thus train using a total of 128 labeled posts per instance per task.\footnote{Note, the choice of 8 training steps is arbitrary, and as with all machine learning problems, we expect the performance to increase with an increase in the number of training steps and the number of labeled posts. However, our objective here is to assess the minimum number of posts a volunteer administrator must annotate to achieve reasonably good performance.} 
We argue that this would be tractable even for low-resource administrators.

\pb{Perturbed Training Set.}
Recall that we use consistent and standardized definition, as determined by the Perspective API, for the harmful content detection task (see \S\ref{sec:dataset}). However, in reality, instance administrators may make mistakes when annotating posts. Further, administrators may exhibit inconsistent views over time. To assess the impact of such annotation inconsistency, we generate a second training set for the harmful content detection task, which we refer to as the \emph{perturbed} training set. Here, we introduce noise into the \emph{original} training set by randomly flipping the labels of 25\% of the posts at each training step. Specifically, at each step, we flip the labels of 4 out of 16 posts and repeat the training process. Note that we still test using the same test set as before. Additionally, we do not generate a perturbed training set for the bot content detection and content warning assignment tasks, as the labels for these tasks are already crowd-sourced and contain inherent variation.

\pb{Classification Model.}
Using the training data, we then fine-tune an (Encoder-only) Transformer Language Model for sequence classification, known for its data-efficient approach to text classification~\cite{sun2019fine}. %
Given the multilingual nature of Mastodon instances, we use Multilingual BERT (mBERT)~\cite{devlin2018bert} (172M parameters) as the base model due to its good cross-lingual knowledge transfers~\cite{pires2019multilingual}, and also because it outperformed other similar sized models in our initial experiments (see Appendix~\ref{sec:appendixinitialexperiments}).
The storage cost of mBERT is 714MB, making it suitable for deployment on Mastodon instances. We use HuggingFace's \textit{transformers} library~\cite{wolf2020transformers} for loading the pre-trained mBERT model, and PyTorch~\cite{paszke2019pytorch} as the underlying differentiation framework. Each model is trained for 2 epochs with a learning rate of 5e$^{-5}$ and a batch size of 16.\footnote{These hyper-parameters are suggested in the original BERT paper~\cite{devlin2018bert} for fine-tuning.}

\begin{figure}[t]
  \centering
  \includegraphics[width=\linewidth]{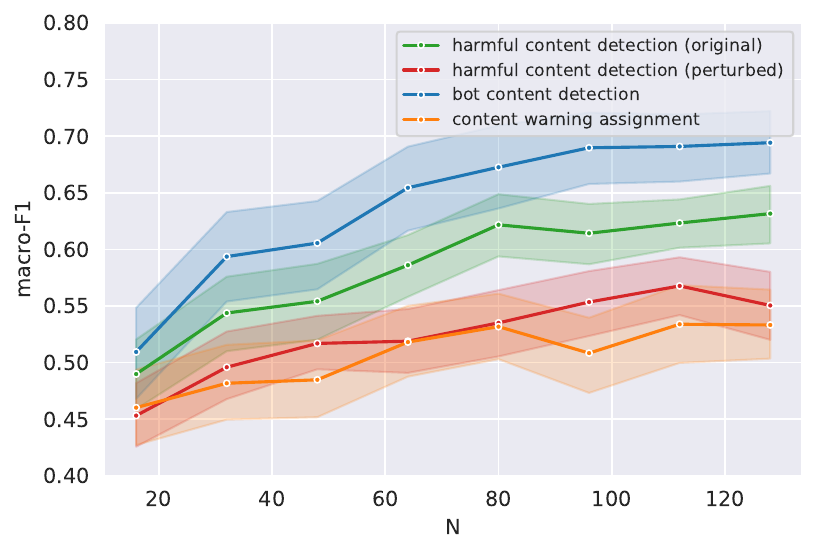}
  \caption{Average macro-F1 scores for local content moderation models across all Mastodon instances after each training step for each content moderation task.} 
  \label{lab:localmoderation}
\end{figure}

\pb{Performance of Local Moderation Models.} We now evaluate the performance of these local content moderation models. Figure~\ref{lab:localmoderation} shows the average macro-F1 scores for local content moderation models across the Mastodon instances for each content moderation task.
To reflect the impact of varying annotation efforts, we plot results after each training step, with $N$ denoting the number of labeled posts used up to each step.
We show results for both the ground truth original training set and the perturbed training set with flipped labels.
To estimate the statistical significance, we apply the Wilcoxon test and list $p$-scores next to the results.
Additionally, we also plot bootstrapped 95\% confidence intervals around the average macro-F1. 

As expected, we observe an increase in the average macro-F1 score as the number of labeled posts ($N$) increases. The local models trained for the bot content detection task achieve the best performance, followed by the models trained for the harmful content detection task. When $N = 128$, the models trained for bot content detection achieve an average macro-F1 score of 0.69, whereas the models trained for harmful content detection using the original training set achieve a modest average macro-F1 score of 0.63. However, the performance of these local models is quite poor for the content warning assignment task, achieving an average macro-F1 score of 0.53 at $N = 128$. This is likely due to significant inter-observer variability, as different users perceive sensitivity differently, leading to uncertainty in labels for the content warning assignment task. This issue is exacerbated when inspecting models for harmful content detection built using the perturbed training set. Here, we observe a notable decrease in the average macro-F1 score across all instances at all values of $N$ compared to using the original training set. At $N = 128$, there is a 12.69\% ($p < 0.001$) drop in the average macro-F1 score when the perturbed training set is employed for each instance.

This underscores the sensitivity of model performance to the accuracy of labels. It also highlights the challenges inherent in local content moderation, particularly in scenarios where there is significant variability in labels due to annotation errors, inconsistent views over time, or differences among administrators.
Finding ways to overcome these challenges is therefore vital.
We argue these problems emerge from a missed opportunity. Instances share posts via federation but do not share content moderation knowledge. Thus, in the next section, we propose a system that can collaboratively train moderation models, implicitly sharing content moderation knowledge across the Fediverse.

\section{F\MakeLowercase{ed}M\MakeLowercase{od} Design}
\label{sec:collaborativecontentmoderation}

To improve automated content moderation and address the challenges associated with local content moderation, we next present \textit{FedMod}, a collaborative content moderation system where instances work together via federated learning.

\subsection{Overview of FedMod Design}

FedMod is a federated learning based moderation system, in which instances exchange information to collaboratively train their moderation models. Here, we focus on classification models for harmful content, bot content, and content warnings. However, we emphasize that FedMod also works for other related classification tasks (\eg spam detection). Within FedMod, instances periodically exchange parameters of partially trained local content moderation models to improve content moderation collectively. 
Our system comprises three iterative steps:
\begin{enumerate} 

 \item [Step 1] \textbf{Model Initialization:} Each instance initially uses its own limited set of labeled posts to train a local content moderation model.

 \item [Step 2] \textbf{Peer Selection:} Each instance then searches for $k$ suitable peers (instances) based on its specific criteria. 

 \item [Step 3] \textbf{Parameter Exchange:} The peer instances exchange parameters of their local models to build a federated community model shared by the instance and its peers. 

\end{enumerate}

These steps are then re-executed at a configurable time interval, $t$, which by default is 24 hours.
FedMod offers the advantage of implicitly sharing linguistic patterns and annotations across instances while safeguarding administrators' privacy by not requiring them to disclose their labeled posts. We detail the full process below.

\pb{Step 1: Model Initialization.}
First, each instance utilizes its limited set of labeled posts to train a local model for a specific moderation task (\eg bot content detection).
We rely on the same model setup as listed in \S\ref{sec:potentialoflocalcontentmoderation}.
Specifically, we fine-tune mBERT~\cite{devlin2018bert} for sequence classification. We use HuggingFace's \textit{transformers} library~\cite{wolf2020transformers} for loading the pre-trained mBERT model, and PyTorch~\cite{paszke2019pytorch} as the underlying differentiation framework.

\begin{figure}[t]
  \centering
  \includegraphics[width=\linewidth]{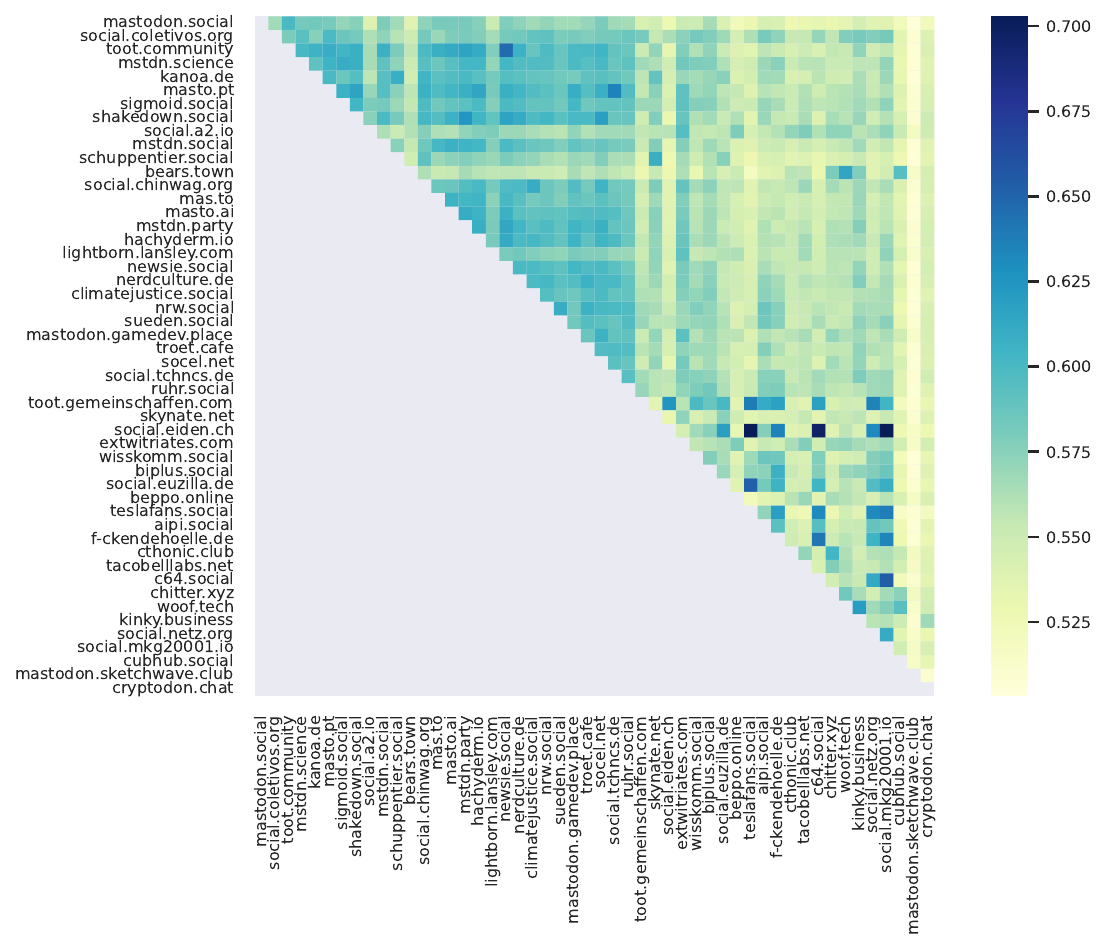}
  \caption{Hashsim similarity across the instances.} 
  \label{lab:heatmap}
\end{figure}

\pb{Step 2: Peer Selection.} Next, each instance must select other suitable peers to partner with. A clear challenge in the FedMod design is identifying such peers. One obvious way for an instance is to randomly pick $k$ other instances as its peers. However, our experience suggests that pairing instances with similar content themes tends to generate better results. 
Thus, we propose an alternative technique, \emph{hashsim}, to identify similar instances. In hashsim, each instance periodically calculates the Jaccard similarity of its own trending hashtags with all other instances and selects the most similar $k$ instances as its peers. 
We use the lightweight Trends API\footnote{<instance.uri>/api/v1/trends/tags} of each instance to retrieve the weekly trending hashtag list. 
To motivate this design, Figure~\ref{lab:heatmap} presents the hashsim similarity across the instances as a heatmap.
Indeed, we see that instance pairs exhibit notable differences.
We emphasize that FedMod's modular design allows for pluggable peer selection strategies or even the use of multiple strategies in parallel.

\pb{Step 3: Parameter Exchange.} Finally, each instance and its selected peers exchange parameters of their partially trained local content moderation models through federated learning~\cite{li2020federated}. Unlike a traditional federated learning setup, there is no central entity in our system to manage the parameter exchange and aggregation. Instead, the Fediverse instances communicate in a peer-to-peer fashion.
After each local training step, each instance requests the local model parameters from its selected $k$ peers. 
Each instance then locally employs \texttt{FederatedAveraging}~\cite{mcmahan2017communication} to create an aggregated model.
The process then repeats after $t$.
We implement parameter exchange using the \texttt{Flower} federated learning framework~\cite{beutel2020flower}. %
\texttt{Flower} provides the infrastructure to do federated learning in an easy, scalable, and privacy-preserving manner. It also supports Differential Privacy (DP) methods~\cite{mcmahan2017learning,andrew2021differentially} thus limiting an individual instance's influence on the aggregated model.

\subsection{Scaling up Peer Selection}
\label{subsec:scalingup}

By default, for the hashsim peer selection, each instance selects peers by calculating the Jaccard similarity of its trending hashtags with all other instances.
While this approach is straightforward, it may lead to scalability issues. For example, just 10,000 instances would require 100 million computations.

To tackle this challenge, we introduce the optional concept of pre-sampling, which involves selecting a small subset of instances for direct similarity calculation. This pre-sampling strategy is based on the observation that instances with a high degree of federation often exhibit high topical similarity.
Thus, each instance locally ranks all other instances it knows by the number of the other instances' users followed by its own users. It then picks the top $i$ instances. 
This estimates the amount of social engagement between users of the two instances. %
Using this, each instance only computes similarity with these $i$ instances. As before, we then select the top $k$ most similar instances from this set of $i$ instances. While this may potentially miss some similar instances, it significantly reduces the number of computations and, consequently, the amount of data exchanged between instances.

\begin{figure*}
     \centering
     \subfloat[][harmful content detection\\ (original)]{
     \includegraphics[width=.25\linewidth]{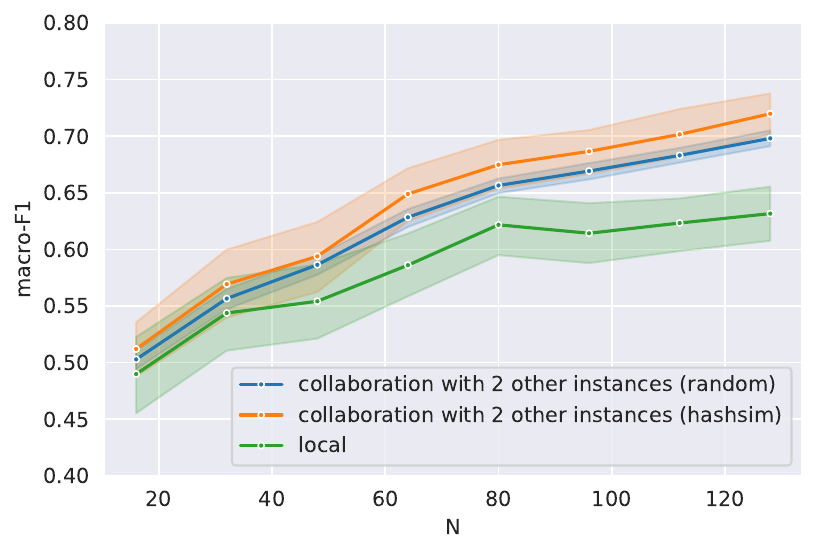} \label{lab:collaborativemoderationa}}
     \subfloat[][harmful content detection\\ (perturbed)]{\includegraphics[width=.25\linewidth]{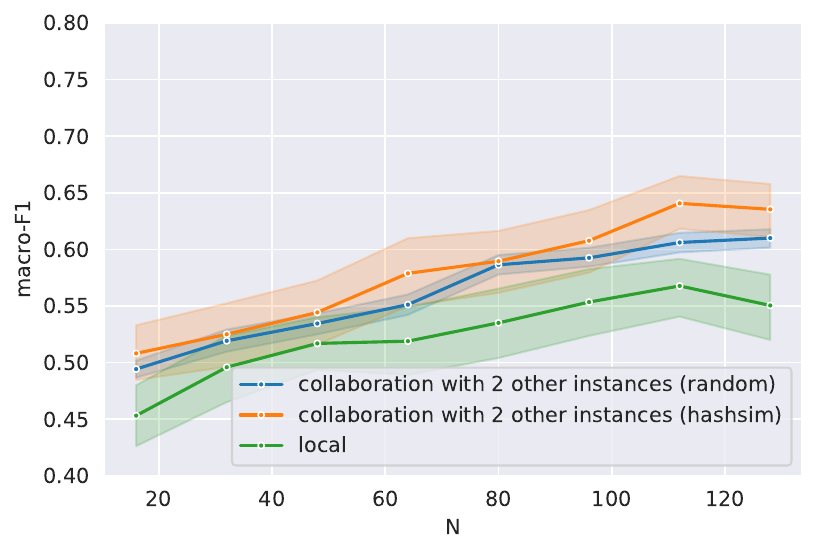}\label{lab:collaborativemoderationb}}
     \subfloat[][bot content detection]{\includegraphics[width=.25\linewidth]{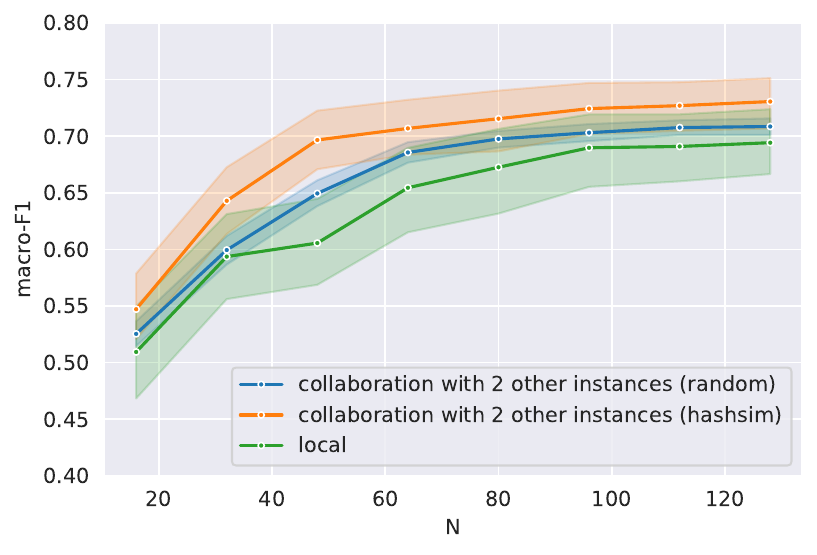}\label{lab:collaborativemoderationc}}
     \subfloat[][content warning assignment]{\includegraphics[width=.25\linewidth]{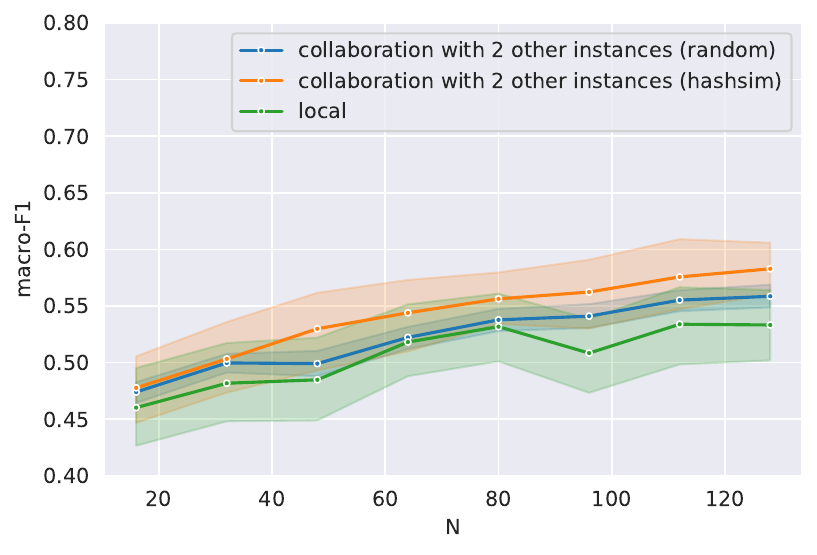}\label{lab:collaborativemoderationd}}
     \caption{Average macro-F1 scores for FedMod based collaborative content moderation models across all Mastodon instances after each training step, using both random and hashsim peer selection for each content moderation task. $N$ denotes the number of labeled posts used up to each training step. %
     }
     \label{lab:collaborativemoderation}
\end{figure*}

\section{F\MakeLowercase{ed}M\MakeLowercase{od} Evaluation}
\label{sec:evaluation}

\begin{figure*}
     \centering
     \subfloat[][harmful content detection (original)]{
     \includegraphics[width=.33\linewidth]{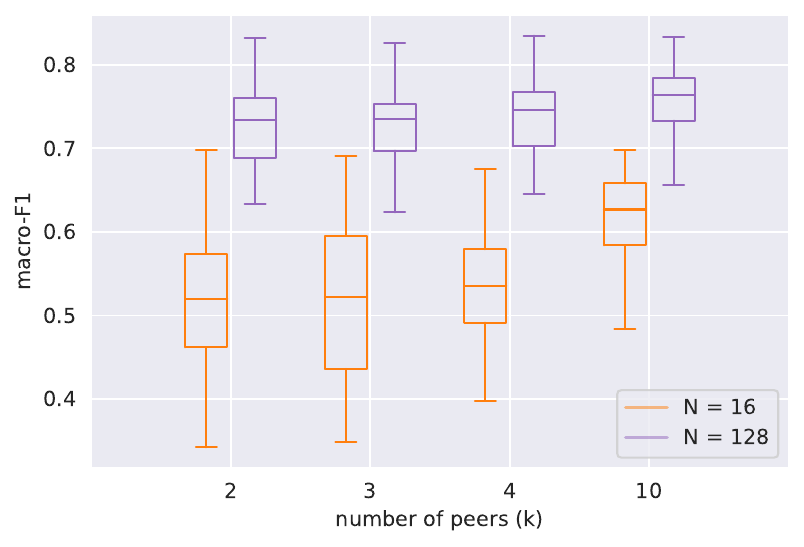} \label{lab:noofpeers}}
     \subfloat[][bot content detection]{\includegraphics[width=.33\linewidth]{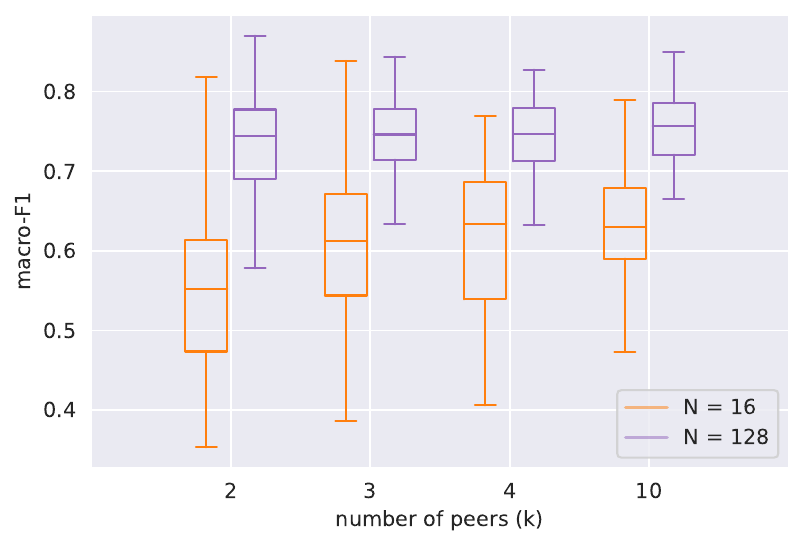}\label{lab:noofpeersa}}
     \subfloat[][content warning assignment]{\includegraphics[width=.33\linewidth]{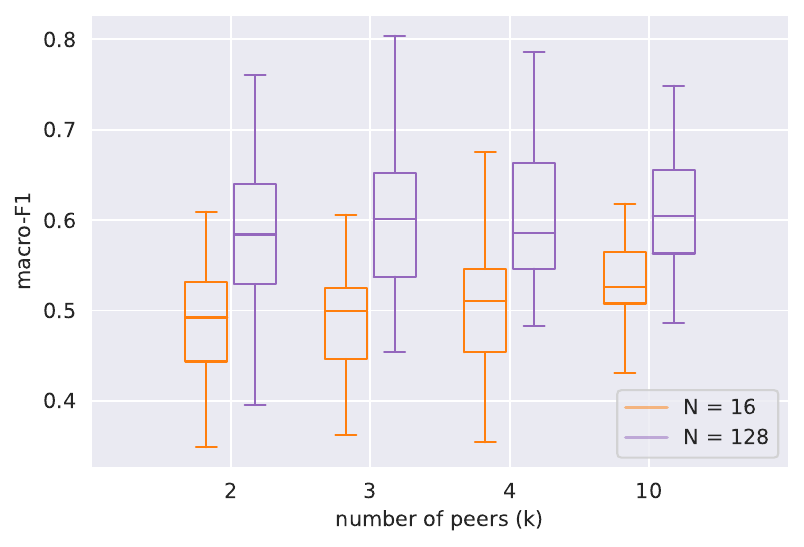}\label{lab:noofpeersb}}
     \caption{Macro-F1 scores of FedMod based collaborative moderation models for each content moderation task across all Mastodon instances, while varying the number of peers ($k$). $N$ denotes the number of labeled posts used up to each training step.
     }
     \label{lab:noofpeersall}
\end{figure*}

To evaluate FedMod, we first focus on its ability to improve upon local content moderation. We then inspect the impact of our pre-sampling strategy and the number of peers.

\subsection{Overall Evaluation}
\label{subsec:overallevaluation}

\pb{Setup.}
We first evaluate FedMod with the number of peers ($k$) set to 2, and then later analyze the impact of increasing $k$ on performance in \S\ref{sec:eval:numpeers}. Initially, we use FedMod without the scalable pre-sampling enabled.
Figure~\ref{lab:collaborativemoderation} presents the average macro-F1 scores across all Mastodon instances for FedMod using both random and hashsim peer selection for each content moderation task.\footnote{Note, in the case of random peer selection, each instance selects $k=2$ other instances 10 times, each time with a different seed. We report the average results.} %
We plot results after each training step, with $N$ denoting the number of labeled posts used up to each step. For the sake of comparability, we also re-plot the results of the local content moderation models, which serve as our baseline (see \S\ref{sec:potentialoflocalcontentmoderation}). Furthermore, we present bootstrapped 95\% confidence intervals around the average macro-F1 scores and apply the Wilcoxon test to assess the significance level of differences in the performance.

\pb{Results.}
FedMod outperforms local content moderation using either of the two peer selection strategies for all three content moderation tasks. %
For instance, for harmful content detection at $N = 128$, FedMod with hashsim achieves an average macro-F1 score that is 12.69\% ($p < 0.001$) better than local content moderation, while FedMod with random peer selection achieves a 9.52\% ($p < 0.001$) improvement. Interestingly, for content warning assignment, where local content moderation performed poorly due to significant variability in labels, FedMod with hashsim achieves an average macro-F1 score that is 9.43\% ($p < 0.001$) better than local content moderation, while FedMod with random peer selection achieves a 3.77\% ($p < 0.01$) improvement. This demonstrates the robustness of collaborative moderation in handling label variability.
Hashsim consistently demonstrates substantially better results compared to random peer selection. For example, at $N = 128$, FedMod with hashsim reaches an average macro-F1 score of 0.73, whereas FedMod with random peer selection achieves 0.70 ($p < 0.05$) for bot content detection. This is likely attributed to hashsim's tendency to find peers that share topical similarities, facilitating the sharing of richer and more relevant linguistic patterns. 

Similar performance patterns are observed when the perturbed training sets are considered for harmful content detection, shown in Figure~\ref{lab:collaborativemoderationb}.
At $N = 128$, FedMod with hashsim significantly outperforms local content moderation by 14.54\% ($p < 0.001$) when a perturbed training set is employed for each instance. This robustness is due to the parameter exchange and averaging inherent in FedMod, which mitigates the impact of individual administrators' mistakes. 
This confirms that FedMod offers an advantage over local content moderation, both in terms of performance and robustness. 
For the remainder of the evaluation, we exclusively utilize FedMod with hashsim due to its better performance compared to random peer selection.

\subsection{Impact of Number of Peers ($k$)} 
\label{sec:eval:numpeers}

\pb{Setup.}
Previously, we evaluated FedMod with $k = 2$, \ie each instance selects two peers for federated parameter exchange.
Next, we analyze the performance impact of the number of peers by varying the value of $k$. Specifically, we re-run the experiments from \S\ref{subsec:overallevaluation}, varying $k$ to 3 and 4, to see how well FedMod adapts to gradual increases in the degree of collaboration. Additionally, we run experiments with $k = 10$ to assess the robustness of FedMod under substantially larger collaborative scenarios.

\pb{Results.}
Figure~\ref{lab:noofpeersall} presents the macro-F1 scores of FedMod based collaborative content moderation models for each content moderation task across all Mastodon instances while varying numbers of peers ($k$). 
For all moderation tasks, we observe a gradual increase in the average macro-F1 score with an increase in the number of peers. However, the performance improvement is much more pronounced when the number of labeled posts contributed by each instance is small. For example, for harmful content detection at $N = 128$, increasing the number of peers from 2 to 10 results in an average macro-F1 improvement of 4.22\% ($p < 0.001$). 
In contrast, at $N = 16$, the same increase in the number of peers results in an average macro-F1 improvement of 19.60\% ($p < 0.0001$). This behavior aligns with intuition: When each instance contributes a small number of labeled posts, the global model, obtained by averaging parameters shared by a large number of peers, captures more diverse patterns than the one obtained with a small number of peers. Thus, we conclude that if the number of labeled posts per instance is small, it is beneficial to engage in collaborative moderation with a larger number of similar instances.

\subsection{Impact of Scalable Pre-Sampling ($i$)}

\begin{figure*}
     \centering
     \subfloat[][harmful content detection (original)]{
     \includegraphics[width=.33\linewidth]{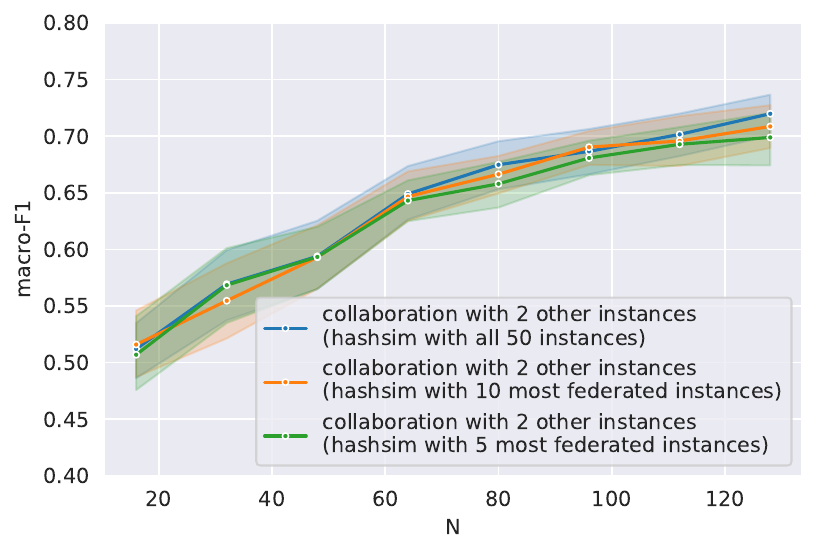} \label{lab:scalability}}
     \subfloat[][bot content detection]{\includegraphics[width=.33\linewidth]{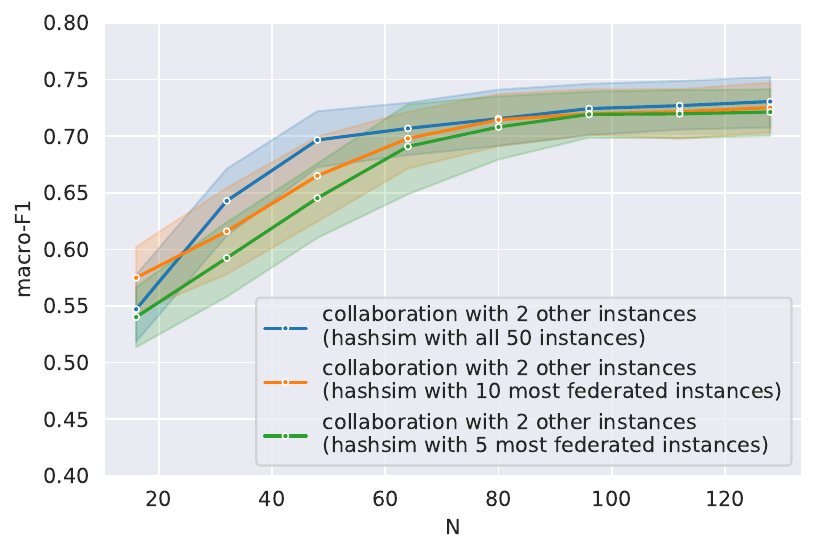}\label{lab:scalabilitya}}
     \subfloat[][content warning assignment]{\includegraphics[width=.33\linewidth]{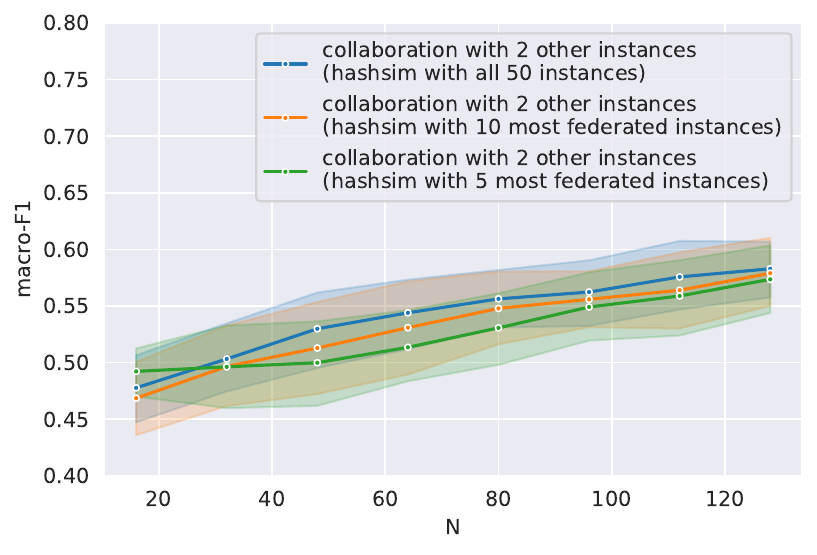}\label{lab:scalabilityb}}
     \caption{Average macro-F1 scores of FedMod based collaborative moderation models for each content moderation task across all Mastodon instances after each training step with and without pre-sampling. $N$ denotes the number of labeled posts used up to each training step.
     }
     \label{lab:scalabilityall}
\end{figure*}

\pb{Setup.}
Recall, we offer an optional scalable pre-sampling of instances to accelerate the identification of suitable instances to peer with.
To assess the performance impact of our scalable pre-sampling approach (\S\ref{subsec:scalingup}), we re-evaluate FedMod using pre-sampling with $i = \{5, 10\}$ and compare the results with the prior approach of calculating similarity with all instances.

\pb{Results.}
Figure~\ref{lab:scalabilityall} presents the average macro-F1 scores of FedMod for each content moderation task across all instances with and without pre-sampling. We plot results after each training step, with $N$ denoting the number of labeled posts used up to each step.
We observe only a slight degradation in average performance when using the pre-sampling strategy compared to when peers are selected from the entire pool of instances. For example, at $N = 128$, when peers are selected from the pre-sampled set of the top 5 most federated instances, the average performance across all instances decreases by only 2.81\% ($p = 0.11$) for harmful content detection, 1.36\% ($p = 0.07$) for bot content detection, and 1.72\% ($p = 0.73$) for content warning assignment. Furthermore, this comes with a 90\% reduction in the number of hashtag similarity computations (and associated data exchange).
This can be attributed to the fact that instances often federate with others who engage in discussions about similar topics, thereby increasing the likelihood of finding similar instances in the pre-sampled set. In fact, 56\% of the instances in our dataset have the most similar instance in their set of the top 5 most federated instances they pre-sampled.

\subsection{Impact of Variation in Moderation Policies}
\label{sec:limitations}

\begin{figure*}
     \centering
     \subfloat[][]{
     \includegraphics[width=.5\linewidth]{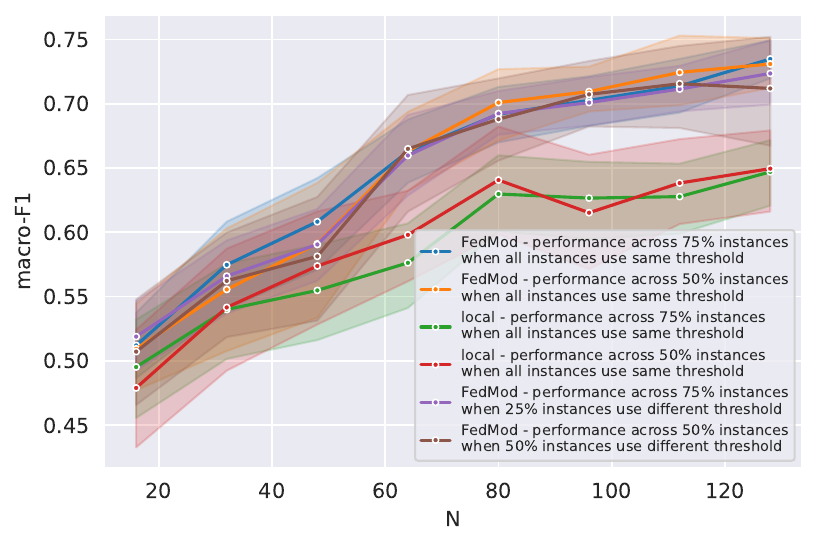} \label{lab:differentthreshold}}
     \subfloat[][]{\includegraphics[width=.5\linewidth]{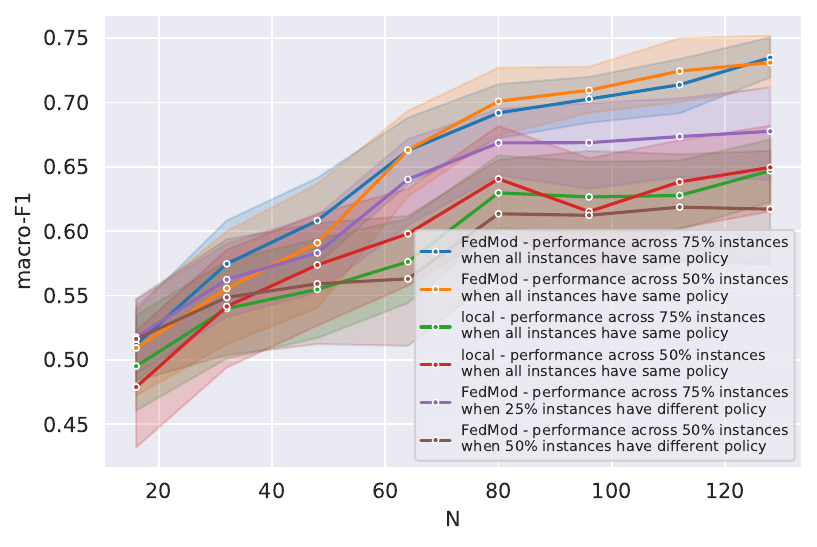}\label{lab:differentpolicy}}
     \caption{Average macro-F1 scores for local and FedMod based collaborative content moderation models across Mastodon instances after each training step (a) with the varying Perspective API thresholds and (b) with the different moderation policies. $N$ denotes the number of labeled posts used up to each training step.
     }
     \label{lab:different}
\end{figure*}

\pb{Setup.}
Recall that different administrators may moderate their instances differently (see \S\ref{subsec:challengesofcontentmoderation}).
To evaluate the impact of these variations on FedMod, we reconfigure our training and testing sets for a subset of instances for harmful content detection task.\footnote{Note, we do not perform this evaluation for the bot content detection or content warning assignment tasks, as the labels for these tasks are already crowd-sourced and contain inherent variation.}
For this subset, we use a higher Perspective threshold (0.8), and only these severely harmful posts will be moderated.
Specifically, we randomly select $x$\% of the instances where $x = \{25, 50\}$ and change their threshold, such that their administrators only label a post as harmful if its Perspective score exceeds 0.80. 
This creates a subset of instances that are more tolerant of harmful posts than the others that retain a threshold of 0.5.
We then proceed to re-evaluate FedMod with the same configurations as detailed in \S\ref{subsec:overallevaluation}. As a baseline comparison, we also rerun local content moderation models for all instances with the same parameters as detailed in \S\ref{sec:potentialoflocalcontentmoderation}.

\pb{Results.}
Figure~\ref{lab:differentthreshold} plots the average macro-F1 scores for the subset of instances that retain a threshold of 0.5. We present FedMod results for both scenarios: \one where x\% of instances employ a 0.8 threshold; and \two when all instances utilize the same 0.5 threshold. For reference, we also re-plot the results of local models when all instances use the same Perspective API threshold. We observe only a slight degradation in FedMod's performance when a subset of instances employ a different Perspective threshold compared to when all instances use the same threshold. 
For example, in the former scenario where 50\% of instances use a 0.8 Perspective threshold, at $N = 128$, the average macro-F1 score for the remaining 50\% of instances that use a 0.5 threshold decreases by just 2.73\% ($p = 0.37$). It is also important to note that 68\% of these 0.5 threshold instances collaborate with at least one instance using a 0.8 threshold. We argue that FedMod maintains robust performance despite a subset of instances using different thresholds because the type of moderated content still overlaps significantly between instances.

\pb{Opposing Moderation Policies.} The above emulates smaller variations in moderation policies. %
Therefore, we next test a hypothetical scenario where instances have completely opposite moderation policies. %
For a subset of x\% of instances where $x = \{25, 50\}$, we completely flip their labels and annotate a post as harmful if the score of any Perspective attribute is below 0.50 (and vice versa).
For the remaining instances, we retain the same definition. We then re-evaluate FedMod and local content moderation, and report results in Figure~\ref{lab:differentpolicy}. 
This tests what happens when instances, who have entirely opposite moderation policies, federate.

Unsurprisingly, we observe a significant drop in the performance of FedMod when 25\% of instances use an entirely opposite definition of harmfulness. For example, at $N = 128$, the average macro-F1 score for the remaining 75\% of instances decreases by 8.21\% ($p < 0.01$). That said, the average FedMod performance for these 75\% of instances is still 4.68\% better than their local models ($p < 0.05$). We observe a further decrease in FedMod performance when the percentage of instances employing opposing policies increases to 50\%, in which case it becomes less effective than local models.
This occurs because, when an instance collaborates with another instance that has completely opposing moderation policies, the resulting federated model fails to learn features that accurately classify opposing views. 
Finding ways to overcome this issue forms an interesting line of future work.

\section{Related Work}

Moderation of user-generated harmful content, such as hate speech \cite{rottger2022data}, offensive language~\cite{zampieri2019predicting,rosenthal2020solid}, anti-Semitic~\cite{warner2012detecting}, misogynous~\cite{pamungkas2020misogyny} and Islamophobic~\cite{vidgen2020detecting} content has remained a key topic of research.
These works mostly attempt to build supervised machine learning classifiers for the automatic identification of harmful content. Davidson \emph{et al.}~\cite{davidson2017automated} train a multi-class logistic regression classifier on a dataset of 25K labeled tweets to distinguish between hate speech, offensive language, and neutral text. Badjatiya \emph{et al.}~\cite{badjatiya2017deep} use deep learning~\cite{lecun2015deep} methods to identify hate speech in text (using 16K annotated tweets). Wulczyn \emph{et al.}~\cite{wulczyn2017ex} use 100K user labeled comments from Wikipedia to identify the nature of online personal attacks. A common aspect of these works is their reliance on a large amount of labeled training data, which renders them impractical for label-scarce scenarios such as the Fediverse.

There exist only a handful of works that discuss content moderation in the Fediverse specifically.
Anaobi \emph{et al.}~\cite{hassan2021exploring} highlight the challenges of instance administrators in the Fediverse and propose a tool to semi-automate policy generation~\cite{anaobi2023will}. Closest to our work is~\cite{bin2022toxicity}, which proposed a system for content moderation on Pleroma. In their system, instances share fully trained local content moderation models with other instances. Their local models are based on a logistic regression classifier with bag-of-words features, requiring thousands of labeled samples to achieve reasonably good performance.
In contrast to FedMod, this places a significant manual burden on administrators. Additionally, their system assumes the existence of high-resource instances capable of training local moderation models, which could then be shared with low-resource instances lacking such resources. 
In contrast, FedMod facilitates the exchange of parameters from partially trained local moderation models among similar peers through federated learning, eliminating the need for reliance on high-resource instances.  

\section{Conclusion \& Future Work}

In the Fediverse, individual instances define their own moderation policies. This makes content moderation challenging, as it is difficult for individual administrators to train their own local models.
To address this problem, we propose FedMod, allowing instance administrators to implement their own moderation policies and collaborate with similar instances to enhance performance and reduce workload. FedMod demonstrates robust performance on three different content moderation tasks: harmful content detection, bot content detection, and content warning assignment, achieving average per-instance macro-F1 scores of 0.71, 0.73, and 0.58, respectively. We also emphasize that our methodology is equally applicable to other forms of content classification, such as spam detection. 

While our design addresses some of the key challenges, other interesting issues remain. Most notably, as demonstrated in \S\ref{sec:limitations}, FedMod's performance degrades when instances with opposing moderation policies collaborate.
These experiments were purposefully unrealistic, and in practice, it is unlikely that (m)any instances would have extreme opposing policies like this. However, it is feasible that malicious actors may purposefully do this to undermine other instances.
In the future, we plan to explore reputation schemes and alternative peer selection strategies that could minimize such risks.
We are also keen to explore more longitudinal aspects of collaborative moderation, \eg how might models evolve over many months of collaboration. 
Additionally, we hope to evaluate FedMod by deploying it among instance administrators. Through this, we hope to gain further evaluative insights into how FedMod can be improved.

\bibliographystyle{ACM-Reference-Format}
\bibliography{main}

\appendix

\section{Appendix}

\begin{table*}[h]
  \caption{Perspective API attributes and their descriptions.}
  \label{tab:perspectiveattibutedescription}
  \begin{tabular}{ll}
    \toprule
    Attribute&Description\\
    \midrule
    \texttt{TOXICITY} & A rude, disrespectful, or unreasonable comment that is likely to make\\&people leave a discussion.\\
    \texttt{SEVERE\_TOXICITY} & A very hateful, aggressive, disrespectful comment or otherwise very\\&likely to make a user leave a discussion.\\
    \texttt{IDENTITY\_ATTACK} & Negative or hateful comments targeting someone because of their identity.\\
    \texttt{INSULT} & Insulting, inflammatory, or negative comment towards a person or\\&a group of people.\\
    \texttt{PROFANITY} & Swear words, curse words, or other obscene or profane language.\\
    \texttt{THREAT} & Describes an intention to inflict pain, injury, or violence against an\\&individual or group.\\
  \bottomrule
\end{tabular}
\end{table*}

\subsection{Ethics}
\label{sec:app:ethics}

We exclusively collect publicly available posts, following well established ethical procedures for social data~\cite{townsend2017ethics}.
We anonymize user identifiers, and make no attempt to link activities to other accounts or real-world identifies.
We use the Perspective API solely to score posts, and no data is stored on the Perspective API server (as we set \texttt{doNotStore} to \texttt{true}).
Data is stored securely within a university silo, and no external access is given.
We do not perform any interactions with users and refrain from scraping full user profiles, followers, or followees. We have obtained ethics approval from the ethics committee at the author's institution. 
We argue that our work can help improve moderation practices in-the-wild, and has key societal benefits.

\subsection{Perspective API Attributes}
\label{sec:appendixdataset}

Perspective is an API that uses machine learning to identify “harmful” posts, making it easier to host better conversations online. Perspective API is the product of a collaborative research effort by Jigsaw and Google’s Counter Abuse Technology team. The API predicts the perceived impact a post may have on a conversation by evaluating that post across a range of emotional concepts, called attributes. Table~\ref{tab:perspectiveattibutedescription} presents the names and descriptions of Perspective API attributes. 

\subsection{Direct use of Perspective API}
\label{sec:directperspective}

While we initially utilize the Perspective API to label Mastodon posts in offline mode to assess the technical feasibility of FedMod, using it directly for online content moderation within the Fediverse presents notable challenges that FedMod avoids:

\begin{itemize}

\item Firstly, the Perspective API provides only a limited set of attributes, potentially misaligning with the nuanced moderation policies unique to individual instances. This limitation hinders the precise enforcement of community-specific guidelines, making it essential to have a more tailored and adaptable solution like FedMod.

\item Secondly, relying solely on the Perspective API would increase the Fediverse's dependence on a centralized service managed by Jigsaw and Google’s Counter Abuse Technology team. Many instance administrators would be unhappy about having to upload all post data to a centralized service. Thus, this contradicts the decentralized ethos of the Fediverse, and it could pose concerns about potential overreliance and vulnerability to external factors beyond administrator control.

\item Thirdly, even though a limited use of the Perspective API is currently provided free of charge, its Terms of Service explicitly state that future tiers may incur charges. Many Fediverse instances are crowdfunded, and these additional costs would be very undesirable.

\end{itemize}

\subsection{Initial Classifier Experiments}
\label{sec:appendixinitialexperiments}

\begin{figure}[t]
  \centering
  \includegraphics[width=\linewidth]{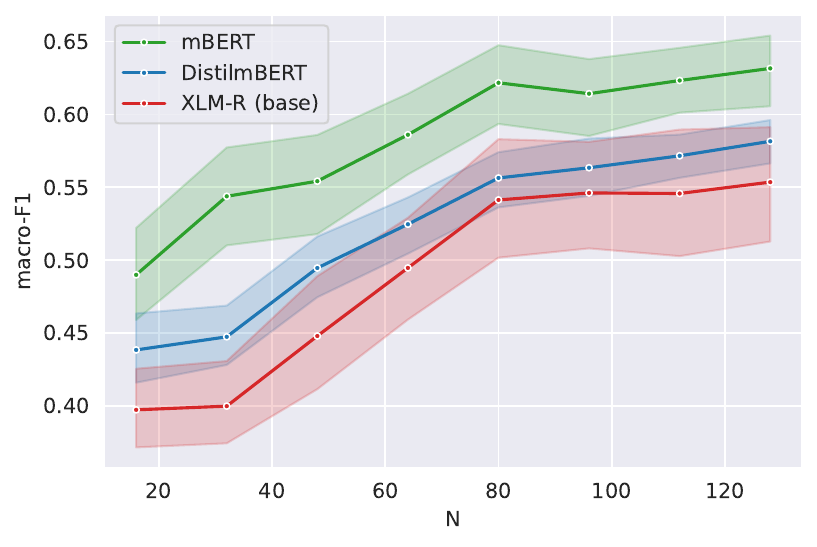}
  \caption{Average macro-F1 scores across all Mastodon instances after each training step using different base language models. $N$ denotes the number of labeled posts used up to each training step.}
  \label{lab:initialexperiments}
\end{figure}

\begin{figure*}[t]
     \centering
     \subfloat[][harmful content detection (original)]{
     \includegraphics[width=.50\linewidth]{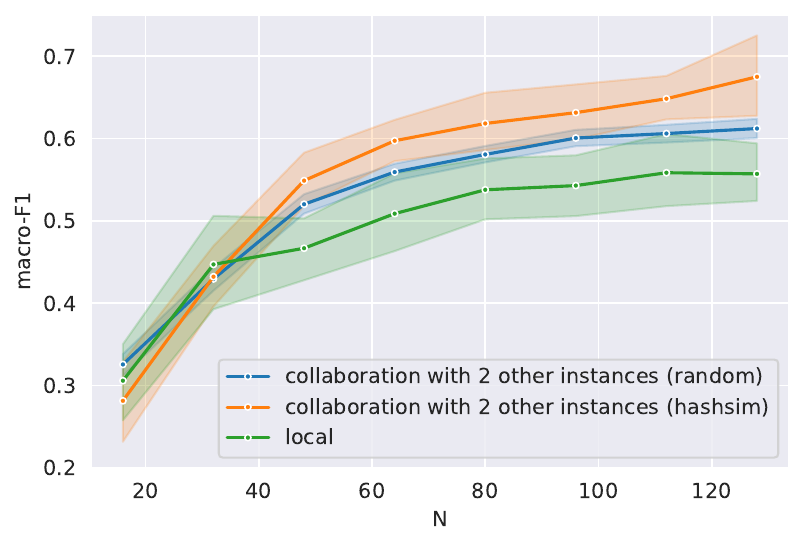} \label{lab:collaborativemoderationastrict}}
     \subfloat[][harmful content detection (perturbed)]{\includegraphics[width=.50\linewidth]{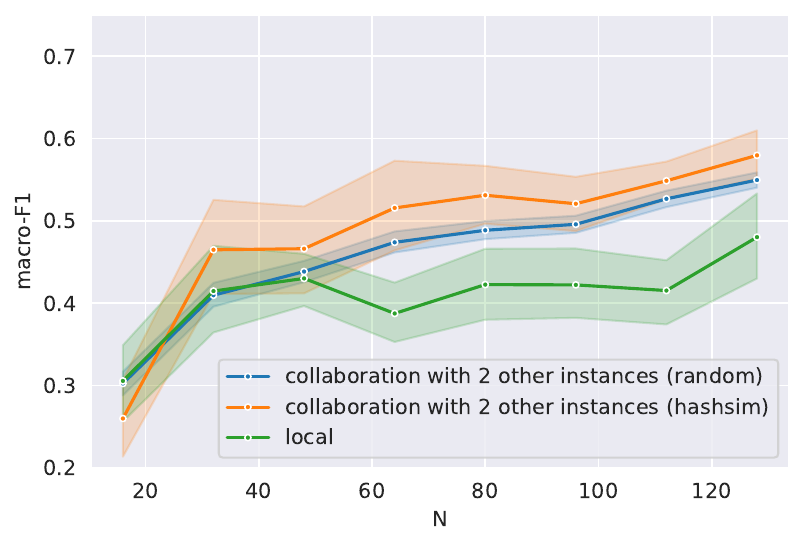}\label{lab:collaborativemoderationbstrict}}
     \caption{Average macro-F1 scores across all Mastodon instances for local content moderation and FedMod (using both random and hashsim peer selection) for harmful content detection. $N$ denotes the number of labeled posts used up to each training step.
     }
     \label{lab:collaborativemoderationstrict}
\end{figure*}

To select a base model for sequence classification, we initially experimented with multiple (Encoder-only) Multilingual Transformer Language Models. Specifically, we fine-tuned mBERT~\cite{devlin2018bert} (172M parameters), DistilmBERT~\cite{sanh2019distilbert} (134M parameters) and XLM-R$_{base}$~\cite{conneau2019unsupervised} (270M parameters) for harmful content detection using original training set of each Mastodon instance. 
Figure~\ref{lab:initialexperiments} presents average macro-F1 scores across all Mastodon instances after each training step, with $N$ denoting the number of labeled posts used up to each step. mBERT consistently outperformed both DistilmBERT and XLM-R$_{base}$. 
For instance, at $N = 128$, mBERT achieves an average macro-F1 score that is 8.62\% ($p < 0.001$) better than DistilmBERT and 14.54\% than XLM-R$_{base}$ ($p < 0.001$). Thus, we use mBERT for our main body of experiments. 

\subsection{Additional Evaluation}
\label{sec:appendixadditionalevaluation}

Previously, we evaluated the harmful content detection on Mastodon instances using 0.5 as a Perspective API threshold. For completeness, we re-plot the results here with a higher threshold of 0.8.

\pb{Setup.} We reconfigure our training and testing sets for all instances using a higher Perspective API threshold of 0.80. Specifically, we refer to a post as \emph{harmful} if its score for any of the Perspective attributes is greater than 0.8, and conversely, as non-harmful if it falls below this threshold. We also regenerate the perturbed training set to assess the impact of annotation inconsistency (see \S\ref{sec:dataset}). We then proceed to re-evaluate local content moderation and FedMod with the same configurations as detailed in \S\ref{sec:potentialoflocalcontentmoderation} and \S\ref{subsec:overallevaluation} respectively. 

\pb{Results.} Figure~\ref{lab:collaborativemoderationstrict} shows the average macro-F1 scores across all Mastodon instances for local content moderation and FedMod (using both random and hashsim peer selection) utilizing both the original (Figure~\ref{lab:collaborativemoderationastrict}) and perturbed (Figure~\ref{lab:collaborativemoderationbstrict}) training sets (now with a Perspcetive threshold of 0.8). We plot results after each training
step, with $N$ denoting the number of labeled posts used up to each step. FedMod outperforms the local content moderation using either of the two peer selection strategies. For instance, at $N = 128$, FedMod with hashsim
achieves an average macro-F1 score that is 21.81\% ($p < 0.001$) better than local content moderation, while FedMod with random peer selection achieves a 10.90\% ($p < 0.001$) improvement. Hashsim consistently demonstrates substantially better results compared to the random peer selection. At $N = 128$, FedMod with hashsim reaches
an average macro-F1 score of 0.67, whereas FedMod with random peer selection achieves 0.61 ($p < 0.05$). 

Similar performance patterns are observed when the perturbed training sets are considered. At $N = 128$, FedMod with hashsim significantly outperforms local content moderation by 18.75\% ($p < 0.001$) when a
perturbed training set is employed for each instance. These results emphasize that FedMod is an effective moderation system irrespective of the labeling methodology. 

\end{document}